\documentclass[prb,twocolumn,superscriptaddress,showpacs]{revtex4-1}

\pdfoutput=1
\usepackage{dcolumn}% Align table columns on decimal point
\usepackage{bm}% bold math
\usepackage{amsfonts}
\usepackage[pdftex]{color,graphicx}
\usepackage{amssymb}
\usepackage{rotate}
\usepackage{subfigure}
\usepackage{subfigure}
\usepackage{amsmath}
\usepackage{amsthm}
\usepackage{hyperref}
\usepackage{times}

\newcommand{\avg}[1]{\left< #1 \right>}

\newcommand{\seq}[1]{\langle #1 \rangle}

\graphicspath{{figs/}}

\begin{document}

% title, abstract (fold)

\title{Strong-disorder renormalization for interacting non-Abelian anyon systems in two dimensions}
\hypersetup{pdftitle={Strong-disorder renormalization for interacting non-Abelian anyon systems in two dimensions},
	pdfauthor={C.R. Laumann}}
	
% authors, date, etc (fold)
\author{C.R. Laumann}
\affiliation{Department of Physics, Harvard University, Cambridge, MA 02138}

\author{D.A. Huse}
\affiliation{Department of Physics, Princeton University, Princeton, NJ 08544}

\author{A.W.W. Ludwig}
\affiliation{Department of Physics, University of California, Santa Barbara, CA 93106}

\author{G. Refael}
\affiliation{Department of Physics, California Institute of Technology, Pasadena, CA 91125}

\author{S. Trebst}
\affiliation{Microsoft Research, Station Q, University of California, Santa Barbara, CA 93106}
\affiliation{Institute for Theoretical Physics, University of Cologne, 50937 Cologne, Germany}

\author{M. Troyer}
\affiliation{Theoretische Physik, ETH Zurich, 8093 Zurich, Switzerland}

\date{\today}

% authors, date, etc (end)

\begin{abstract}
We consider the effect of quenched spatial disorder on systems of interacting, pinned non-Abelian anyons 
as might arise in disordered Hall samples at filling fractions $\nu=5/2$ or $\nu= 12/5$. 
In one spatial dimension, such disordered anyon models have previously been shown to exhibit a hierarchy 
of infinite randomness phases. 
Here, we address systems in two spatial dimensions and report on the behavior of Ising and Fibonacci anyons under the numerical strong-disorder renormalization group (SDRG). 
In order to manage the topology-dependent interactions generated during the flow, we introduce a \emph{planar} approximation to the SDRG treatment. We characterize this planar approximation by studying the flow of disordered hard-core bosons and the transverse field Ising model, where it successfully reproduces the known infinite randomness critical point with exponent $\psi \approx 0.43$. 
Our main conclusion for disordered anyon models in two spatial dimensions is that systems of Ising anyons as well as systems of Fibonacci anyons do not realize infinite randomness phases, but
flow back to {\em weaker} disorder under the numerical SDRG treatment. 
\end{abstract}


% 73.43.-f: Quantum Hall effects
% 75.10.Nr: Spin-glass and other random models
% 72.15.Rn: Localization effects (Anderson or weak localization)
% 03.65.Vf: Topological phases (quantum mechanics)

\maketitle

%\tableofcontents

% title, abstract (end)

\section{Introduction} % (fold)
\label{sec:introduction}

Since von Klitzing's seminal discovery of the quantized Hall effect \cite{Klitzing}, non-symmetry breaking topological order has become an essential part of our understanding of low-temperature electronic systems.\cite{wen2004quantum}
The fundamental feature of many of these topological phases of matter is the presence of anyonic quasiparticles, whose adiabatic exchange entails a nontrivial operation on the state of the system in contrast to the signs accumulated by conventional fermions and bosons. 
In the most general case, the exchange of such anyons induces non-Abelian transformations on a (topologically) degenerate manifold of states of the system.
There are several candidate systems currently under intense experimental scrutiny, which on theoretical grounds have been proposed to exhibit the simplest incarnation of such non-Abelian quasiparticles, so-called Ising anyons. These include the Moore-Read state \cite{Moore1991362} proposed for the fractional quantum Hall liquid at filling fraction $\nu=5/2$, p$_x$+ip$_y$ superconductors,\cite{Read:2000vz} heterostructures involving topological band insulators \cite{Fu:2008gu,Sau:2010cl,*Alicea:2010hy} and certain frustrated magnets.\cite{Chaloupka:2010gi,*Jiang:2011hx,*Singh:2011uv}
An incarnation of slightly more complicated anyons, so-called Fibonacci anyons, have been proposed in a theoretical description of the FQH state at filling $\nu=12/5$ based on the Read-Rezayi state.\cite{Read:1999wx}
In real samples, however, unavoidable impurities pin these particles randomly in space and residual microscopic interactions split the various fusion channels for the associated topological charge. The collective behavior due to this disorder pinning may strongly influence thermodynamic transport and topological interferometry experiments. It may also impede the use of such systems as quantum computers in proposed schemes of topological quantum computation\cite{Kitaev:2003fk,*Nayak:2008dp}. 
%\\

The pinned anyon problem in a disordered Hall bar is reminiscent of a quantum spin glass~\cite{Laumann:2011vn}. 
However, within the Strong Disorder Renormalization Group (SDRG) method\cite{PhysRevLett.43.1434,*PhysRevB.22.1305} that we use in this article, the non-Abelian character of the fusion rules imply that the interactions generated by the SDRG are intrinsically topology-dependent. 
The action of this renormalization group (RG) on fusion in one piece of the sample may influence the fusion of particles, and thus their interactions, elsewhere in the sample, even if they are a priori not connected by direct energetic interactions. 
Hence, many of the properties of the SDRG known from applications to traditional spin systems, which have simple tensor product Hilbert spaces, must be revisited for non-Abelian systems. 
The {\it one-dimensional} incarnations of these interacting anyon problems have been solved both in the pure\cite{PhysRevLett.98.160409} and, within the SDRG, in the disordered\cite{Bonesteel:2007,PhysRevB.78.224204,PhysRevB.79.155120} case. In the latter case, the SDRG analysis found flows to a hierarchy of infinite randomness fixed points whose specific character depends on the underlying anyon theory. 
In this report, we use the SDRG to numerically study disordered non-Abelian anyon systems in {\it two spatial dimensions}. We use several different approximations to handle the explicitly topology-dependent interactions generated by the SDRG renormalization scheme. Our main conclusion from such an SDRG treatment is that in the presence of disorder two-dimensional systems of interacting Ising or Fibonacci anyons do {\it not} realize infinite randomness phases, but that these systems flow back to {\it weaker disorder} under the SDRG.
\footnote{Note that this does not exclude the possibility of a disorder-dominated phase. Our result implies that, within our approximations, no RG-flow to a strong disorder fixed point occurs.}

In the absence of disorder, these 2-D interacting anyon models have been studied previously, and argued to exhibit gapped topological phases on their own\cite{PhysRevB.73.201303,PhysRevLett.103.070401,Ludwig:2011}. Coupled with the results of the present study, one might suspect that quenched disorder could simply be irrelevant (in the RG sense) for two-dimensional anyon models, with all RG flows returning to the clean fixed points. Two observations however suggest that such a scenario cannot represent the complete picture. First, in one dimension, both pure and disordered anyon chains exhibit gapless critical phases which are distinct from the clean systems and are protected by the topological nature of the Hilbert space. Second, by recasting the disordered two-dimensional {\it Ising anyon} problem in terms of non-interacting Majorana fermion zero-modes, some of us\cite{Laumann:2011vn} have recently established the presence of a disorder-induced (but {\it not} infinite randomness) thermal metal phase in the phase diagram of the pinned disordered 2-D {\it Ising anyon} problem. Whether an analog of this thermally conducting 2-D phase might also be found  for the more complicated Fibonacci anyons
(e.g. in the $\nu=12/5$ quantum  Hall state) - or any higher level
anyon model - is an intriguing open question.

In this report, we first briefly review the physical picture of anyons pinned within a two-dimensional quantum Hall sample and motivate the effective Hamiltonian describing their interactions in Sec. II. In the following section III, we discuss the ingredients of a fully two-dimensional strong disorder renormalization group analysis and introduce the {\it planar} approximation, which we invoke to study the infinite randomness behavior of the Fibonacci anyons.
We characterize the planar SDRG by comparing its behavior on the well studied random transverse field Ising model (TFIM) and hard-core boson hopping problem and uncover a pathology which we argue reflects the flow of the more exotic topological models to weaker disorder. We summarize these arguments at the end of Sec.~\ref{sec:strong_disorder_rg}. 
Section~\ref{sec:models} provides a more detailed introduction to each of the models, the strong-disorder rules and a brief summary of the numerical results. Section~\ref{sec:implementation} discusses a few of the implementation details for our numerical study.

\section{Disordered Pinned Anyon Models} % (fold)
\label{sec:pinned_anyons}

In this section, we provide a heuristic introduction to the physics underlying disordered pinned anyon models using the $\nu=5/2$ state as an example. 
For more details, see Refs.~\onlinecite{PhysRevLett.98.160409,Trebst:2008hl,Bonderson:2007uh}.

Let us consider a Hall bar in a strong magnetic field at filling factor $\nu=5/2$ and assume that the fractional quantum Hall phase associated with this filling is indeed described by the Moore-Read Pfaffian state.\cite{Moore1991362} 
In an ideal sample tuned precisely to $\nu=5/2$, the system forms a uniform electron fluid with a gap to quasiparticle (qp) excitations with electromagnetic charge $e/4$ and non-Abelian braiding statistics. These quasiparticles are also 
%called Ising anyons or sigmas, in varying contexts. 
called Ising anyons or ``sigmas'', in varying contexts. 

\begin{figure*}[htbp]
	\centering
		\includegraphics[scale=1]{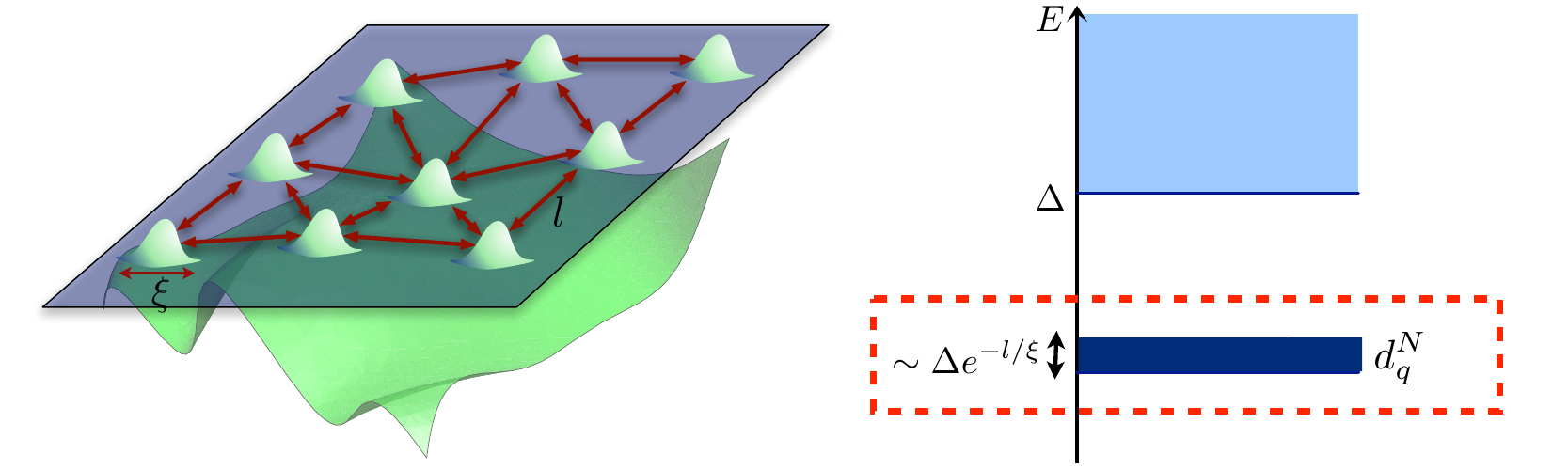}
	\caption{(left) Schematic depiction of quasiparticles of size $\xi$ localized on a randomly displaced triangular lattice within a two-dimensional  sample. (right) Schematic spectrum with a bulk gap $\Delta$ and a topological degeneracy split at an exponentially smaller scale in the interparticle separation $l$.}
	\label{fig:fqh-spec}
\end{figure*}

Detuning the filling fraction away from the center of the $5/2$ plateau or the introduction of a random electrostatic background (eg. due to sample impurities) introduces a finite density of qps into the ground state of the system. In the pure, detuned system, the dilute gas of charged quasiparticles Wigner crystalizes into a triangular lattice; in a weakly disordered sample, the lattice sites randomly shift toward wells of the potential. In either case, the orbital (charge) degree of freedom of these particles ought to gap out of the low energy spectrum, see Figure~\ref{fig:fqh-spec}.

Due to the non-Abelian statistics of the quasiparticles, however, this is not the end of the low energy description of the system. Assuming the $N$ pinned qps are sufficiently far apart, then there remains a manifold of (nearly) degenerate ground states that grows exponentially with the number of qps $N$. 
This degeneracy is in many ways analogous to that of a system of noninteracting spin-1/2 quasiparticles, which would provide the ground state a $2^N$ spin degeneracy. For the non-Abelian qps, the degeneracy depends on the underlying anyonic theory and the so-called quantum dimension of the non-Abelian degree of freedom in this theory: for Ising anyons it asymptotically grows as $\sqrt{2}^N$, while for Fibonacci anyons it grows as $\phi^N$ where $\phi$ is the golden ratio $\phi = (\sqrt{5}+1)/2$ (hence the name Fibonacci anyon). These degeneracies are not associated with any local observables, but rather with the braiding history of the qps.%

We can understand the construction of the topological Hilbert space from its fusion rules. For the $\nu=5/2$ phase, there are three topological charges: the vacuum $1$, the elementary quasiparticle $\sigma$ and the fermion $\psi$. That is, measuring the net topological charge of any collection of quasiparticles will produce one of these three results. 
The topological charge of a single qp is $\sigma$. 

\begin{figure}[htbp]
	\centering
		\includegraphics[width=\columnwidth]{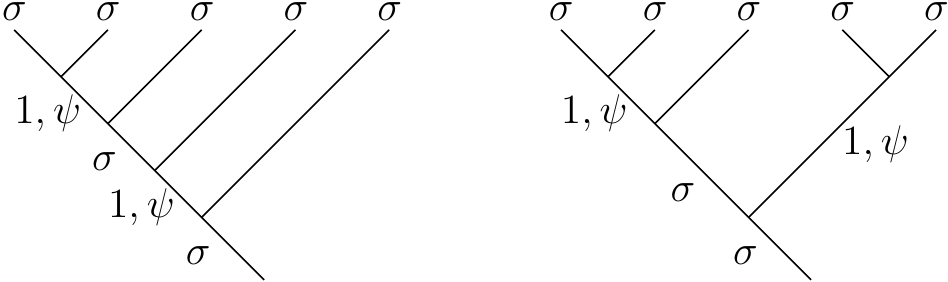}
	\caption{(left) Fusion tree for 5 $\sigma$ particles with all possible labelings of each intermediate fusion channel, consistent with the fusion of the charges from top to bottom. The four available labelings provide a basis for the Hilbert space. (right) Alternative fusion ordering providing a different basis for the same space. The bases may be related by a unitary transformation given by F-moves and braid moves (exchanges).}
	\label{fig:fusion-tree}
\end{figure}

Thus, we can build up a basis for the topological Hilbert space by considering the state space built by successively fusing together each of the $N$ $\sigma$'s in the system. The fusion rule for a pair of $\sigma$ particles is
\begin{eqnarray}
	\label{eq:sigma_fusion_rule}
	\sigma \times \sigma &=& 1 + \psi \,,
\end{eqnarray}
while both the $1$ and $\psi$ particles act like the identity when they fuse with a $\sigma$. An orthogonal basis for the Hilbert space is therefore given by the labelings of a fusion tree (see Fig.~\ref{fig:fusion-tree}) that are consistent with the above fusion rules. Notice that this description of the Hilbert space requires choosing an ordering of the quasiparticles 
which is implicit in the depiction in the form of a fusion tree in Fig.  \ref{fig:fusion-tree}

-- other fusion orderings provide alternative bases which may be related by unitary transformations built out of so-called `F-moves' illustrated in Fig.~\ref{fig:fusion-tree} and braid moves (exchanges). 

The above description of the topological Hilbert space 
corresponds to
 a collection of qps that are arbitrarily far apart -- at finite separations, the degenerate manifold of states will split in some non-trivial way. 
This effect can already be seen for a pair of anyons where topological charge tunneling \cite{PhysRevLett.103.110403} will result in a splitting of the two possible fusion outcomes \eqref{eq:sigma_fusion_rule} for a pair of sigmas. Generalizing this pair-wise splitting into an effective many-body Hamiltonian we arrive at an anyonic version~\cite{PhysRevLett.98.160409} of the conventional Heisenberg model given by
\begin{equation}
	\label{eq:ham_lattice}
	H = \sum_{\seq{ij}} J_{ij} \Pi^{\psi}_{ij} \,,
\end{equation}
where $\Pi^{\psi}_{ij}$ projects onto the $\psi$ fusion channel of the $ij$'th pair of $\sigma$'s and $J_{ij}$ may take any sign and strength depending on which channel is locally favored and by how much. The tunneling-mediated exchange couplings $J_{ij}$ fall off exponentially with the interparticle separation $l_{ij}$ relative to the length scale of the qp wavefunction -- roughly the magnetic length.\cite{PhysRevLett.103.076801} Thus, all of the physics described herein ultimately lies in the narrow band below the gap indicated in Fig.~\ref{fig:fqh-spec}.

It is difficult to get a solid handle on the microscopic physics governing $J(l)$.
Nonetheless, calculations within several theoretical frameworks have been done.
\cite{PhysRevLett.103.110403,PhysRevLett.103.076801,PhysRevLett.103.107001,PhysRevB.79.134515} 
They all find an exponentially decaying envelope expected for a tunneling mediated process in a gapped quantum liquid, 
within which the favored fusion channel and thus the sign of the $J(l)$ oscillate -- akin to an RKKY interaction. 
Thus, unless the system is an
ideal triangular lattice with all $l_{ij}$ equal, one expects that the
$J_{ij}$ indeed have a very broad distribution \emph{and} strong sign
disorder. It is reasonable to expect the physics to be captured by passing to
a disordered ensemble of anyon lattice Hamiltonians $H$ with 
independent identically distributed $J_{ij}$, \`a la the Edwards-Anderson approach to disordered magnets. 
This is also what motivated us to look into strong-disorder approaches 
for a description of the system.

Anyon projector Hamiltonians like \eqref{eq:ham_lattice} depend on geometry in a rather subtle way, unfamiliar from spin models. Indeed, the notation $\Pi^{\psi}_{ij}$ is ambiguous -- in general, the pair-wise interaction of two distinct anyons $i$ and $j$ should be labeled by a path connecting the two anyons which indicates on which side of the other anyons in the system the interaction is mediated. From the point of view of the fusion basis described above, this corresponds to fixing the ambiguity about how to change the fusion ordering in order to implement the interaction of two particles not adjacent in the original ordering. Within the $k=2$ (Ising) anyon theory which we have discussed so far, the non-locality does not play an important role in our further analysis, but for higher $k$, it can 
lead 
apparently uncoupled clusters of anyons
 to generate explicit couplings perturbatively. We will return to this feature in the discussion of the Fibonacci rules in Sec.~\ref{sub:fibonacci_anyons}.

% section pinned_anyons (end)

\section{Strong Disorder Treatment} % (fold)
\label{sec:strong_disorder_rg}

We consider the behavior of pinned anyon lattice models in the strongly disordered regime. In one spatial dimension, such models exhibit a hierarchy of infinite randomness phases under the strong-disorder renormalization group (SDRG) for $SU(2)_k$ anyons indexed by the level $k$.\cite{PhysRevB.79.155120} It is natural to ask whether such infinite randomness behavior carries over to  the two dimensional pinned lattice models that motivated their study. This would be an especially intriguing result as very few two dimensional models are known to flow to such infinite randomness fixed points. In fact there are only two known examples: the transverse field Ising model (TFIM) with random fields and random bonds, which has an infinite randomness critical point (IRCP) separating random ferromagnetic and paramagnetic phases \cite{Motrunich:2000qf}; and the bipartite Majorana hopping model, which has a marginally stable infinite randomness phase.\cite{Motrunich:2002hz} Meanwhile, such simple models as hard-core boson hopping flow away from infinite randomness towards weaker disorder in two dimensions.

In one dimension, many infinite randomness fixed points (IRFPs) may be found analytically because the strong-disorder rules preserve chain geometry and introduce only trivial correlations into the coupling distributions.\cite{Fisher:1992rt,*Fisher:1994p7999,*Fisher:1995vn} This holds even for anyonic chains because fusion need not significantly reorder the anyons in the chain.\cite{PhysRevB.78.224204,PhysRevB.79.155120}  
In two dimensions, however, two complications arise: first, the RG rules for any model generate lots of next-neighbor bonds which quickly render the geometry of an initial 2-D lattice unrecognizable. Moreover, these bonds have significant correlations between their strength and geometric significance, even as the geometry becomes more obscure. Second, for the anyon models at $k>2$, the topology of bonds which \emph{cross} becomes important for the generation of renormalized bonds. As the mesh renormalizes, % into a mess, 
keeping track of such crossings becomes more and more problematic. 

The loss of the lattice geometry renders direct analytic treatments in two dimensions intractable but may be dealt with by numerical simulation of the SDRG flow on large instances. In this approach, we specify the perturbative rules for integrating out strong bonds in a particular model and then iteratively apply these rules to decimate large samples while monitoring the flow of their couplings and geometry. In practice, we also need to control the growth of memory requirements by some truncation of generated weak couplings. This RG predicts its own success or failure: if after an initial transient, we discover that the width of the distribution of log couplings $P_{t}(\ln (\Omega/J))$ falls onto a scaling distribution with a width $w \sim N^{-\psi/ d}$, 
then we have found an IRFP with scaling exponent $\psi$ (here, $N$ is the number of sites remaining in the system and $d$ the spatial dimension).
 If on the other hand, the width $w$ shrinks, then the system is trying to flow back to weaker disorder and the strong-disorder approach is suspect.

The effect of crossings on the interaction renormalization, which arises for  Fibonacci anyons ($k=3$), is much more problematic for a numerical treatment. Rather than  attempt to keep track of the topology of crossing bonds, we invoke the following approximation: since the original lattice is 2-D and we expect the physics to be dominated by nearest neighbor interactions, we require all renormalized interactions to be \emph{planar}. Thus, on a given renormalization step, newly generated bonds are added into the model in order from strongest to weakest so long as they do not make the lattice non-planar. Clearly, this truncation rule is an approximation to the SDRG which is not itself perturbatively controlled. We therefore compare the behavior of the SDRG with and without this planar approximation on a number of (non-)topological models in order to determine if it deems reliable for the case of Fibonacci anyons. 

\begin{table*} % SDRG Models
	\label{tab:sdrg_models}
	\begin{tabular}{l@{ }|@{ }l@{ }|@{ }l@{ }|@{ }l}
	Model			& SDRG Rule			& Bonds  			 						& Asymptotic Flow	\\
	\hline \hline
	Hardcore bosons 		& Planar 		& Long bond disease							& ``marginal'' \\
	       		& Nonplanar	& $\avg{l} \sim N^{-1/2}$ & Steep Weakening     \\
	\hline
	TFIM  			& Planar			& $\avg{l} \sim N^{-1/2}$ & Strengthening, $\psi \approx 0.43$ \\
				& Nonplanar   & $\avg{l} \sim N^{-1/2}$ & Strengthening, $\psi \approx 0.42 \pm 0.06$\footnote{From Ref.~\onlinecite{Motrunich:2000qf}.}\\
	\hline 
	Ising anyons     & Planar		& Falls apart						  & Nonexistent \\
	(Nonbipartite) & Nonplanar & $\avg{l} \sim N^{-1/2}$ & Weakening          \\
	\hline
	Ising anyons   &	Planar    & Falls apart             & Nonexistent \\
	(Bipartite)  & Nonplanar & $\avg{l} \sim N^{-1/2}$ & Shallow weakening  \\
	\hline
	Fibonacci anyons & Planar		& Long bond disease							& ``marginal'' \\
	\end{tabular}
	\caption{Behavior of various models under numerical SDRG in two dimensions. }
\end{table*}

Table~I %\ref{tab:sdrg_models} 
summarizes the qualitative behavior of each of the models that we have studied using our numerical implementation of the SDRG. In order of increasing complexity, these are hard-core boson hopping (aka. XX model), Ising anyons interacting on bipartite and non-bipartite lattices, the transverse field Ising model and the pinned Fibonacci anyon model.  For all of these models except the Fibonacci anyons, we have studied the SDRG flow with the planar approximation and without in order to understand better the behavior of the approximation. The background, RG rules, and quantitative results for each of the models appear in more detail in Sec.~\ref{sec:models}. General implementation details will  shortly be mentioned in Sec.~\ref{sec:implementation}.

The primary results of our investigation into the planar approximation are as follows: 
\begin{itemize}
	\item The planar approximation appears to be asymptotically valid for the one two-dimensional model that is known to flow strongly to infinite randomness --- the TFIM at its infinite randomness critical point. In particular, the existence and scaling properties of the critical point are identical with and without the approximation to within numerical error.
	
	\item The models which flow back to weaker disorder under the usual SDRG are `stabilized' by the planar truncations, in the sense that they find marginal scaling fixed points for their bond strength distributions that are independent of system size. In particular, this behavior can be seen in the hard-core boson model.
	
	\item These ``marginal'' fixed points are, however, physically spurious. At these fixed points, the distribution of strong bond lengths, as measured by the bare geometry of the original lattice positions, saturates the system -- indeed, the strong bond length distribution becomes consistent with a random graph dropped on top of a toroidal geometry. These bare lengths $l$ should scale with the inverse square root of the remaining density, as they do, for example, at the TFIM fixed point. 
\end{itemize}
We take the `long bond disease' exhibited by the bosons and Fibonacci anyons to indicate that the marginal fixed point behavior they exhibit is actually caused by the planar truncation. Indeed, models which flow to weaker disorder generate many strong bonds at each step of the RG, of which many will be cut by the planar truncation rule, falsely preventing the flow back to weaker disorder. 

Finally, we note that the Ising anyon models have their own pathology within the planar approximation: between the opposite sublattice rule and planarity, these models generate so few bonds that they fall apart into disjoint chunks at a finite time in the SDRG flow. This is also a physically spurious result. Indeed, even the numerically observed narrowing of the bipartite Ising anyon coupling distribution without the planar approximation is believed to be a transient in what should be a marginal flow to infinite randomness.\cite{Motrunich:2002hz}

\section{Models} % (fold)
\label{sec:models}

For all of the models we consider, we use the following notational conventions. Latin coupling constants ($J$, $h$) refer to energetic coupling constants in the model and $\Omega$ denotes the current maximum energy coupling. Greek letters  ($\beta$, $\zeta$) refer to log-couplings relative to $\Omega$. $\Omega_0$ is the initial (unrenormalized) strongest coupling and thus $\Gamma = \log \Omega_0 / \Omega$ measures the log energy scale of the flow. The number of sites remaining in the system is given by $N$. $\Gamma$, $\Omega$ and the various coupling distributions generally depend on the time in the renormalization flow $t$; we will generally omit writing this dependence except where necessary.

\subsection{Transverse Field Ising Model} % (fold)
\label{sub:transverse_field_ising_model}

The strong disorder regime of the transverse field Ising model (TFIM) has been studied in great detail both analytically in one dimension\cite{Fisher:1995vn} and numerically in higher dimensions.\cite{Pich:1998ba,Motrunich:2000qf,Kovacs:2010bf,Kovacs:2011ef} In particular, Motrunich \emph{et al.}\cite{Motrunich:2000qf} demonstrated that in two dimensions, the TFIM flows to an infinite randomness critical point separating random quantum ferromagnetic and paramagnetic phases -- as is known analytically in one dimension. The TFIM is thus the fiduciary starting point for our treatment of 2-D models: it provides a check for both our numerical implementation and the validity of the planar approximation. If we are to have any faith in the planar approximation in the more complicated topological models below, it must reproduce the universal properties of the TFIM.

\subsubsection{Model} % (fold)
\label{ssub:tfim_model}

The transverse field Ising model consists of a lattice of spin-1/2 degrees of freedom $\sigma_i$ governed by the Hamiltonian
\begin{equation}
	\label{eq:tfim_ham}
	H = - \sum_{\seq{ij}} J_{ij} \sigma^z_i \sigma^z_j - \sum_i h_i \sigma^x_i - H \sum_i \mu_i \sigma^z_i
\end{equation}
Here the $J_{ij}$ are Ising couplings, $h_i$ are local transverse fields and $H$ is an externally applied uniform field in the $z$ direction that couples to the moments $\mu_i$ of the spins. In the strong disorder regime, the $J_{ij}$ and $h_i$ are broadly distributed random variables whose signs are irrelevant to the thermodynamics -- any frustrated loop will have an extremely weak bond that may be dropped to unfrustrate the system (although clearly this effects the growth of net moments). In this regime it is natural to introduce the logarithmic variables:
\begin{eqnarray*}
	\beta_i &=& \log \Omega / h_i \\
	\zeta_{ij} &=& \log \Omega / J_{ij}
\end{eqnarray*}
where $\Omega$ is the maximum coupling in the system at a given time in the flow.

This system has two kinds of strong-disorder RG rules depending on whether the strongest coupling remaining in the system at scale $\Gamma = \log \Omega_0/\Omega$ is a transverse field $h_i$ or a  bond $J_{ij}$:
\begin{itemize}
	\item Field decimation: the field $h_i$ pins the spin $\sigma_i$ in the $x$ direction so the site $i$ may be removed from the lattice. For each pair $j,k$ of $\sigma_i$'s former neighbors, perturbation theory generates an effective interaction by virtual flipping of $\sigma_i$:
	\begin{equation}
			\label{eq:tfim_rule_fielddec}
			J'_{jk} = J_{jk} + \frac{J_{ji}J_{ik}}{2 h_i}
	\end{equation}
or, in terms of the logarithmic variables:
	\begin{equation}
		\zeta'_{jk} = \min\left\{ \zeta_{jk}, \zeta_{ji} + \zeta_{ik} - \beta_i\right\}
	\end{equation}
	
	\item Bond decimation: the spins $\sigma_i, \sigma_j$ coupled by the strong bond $J_{ij}$ bind together as a single larger Ising moment:
	\begin{equation}
		\label{eq:tfim_rule_bonddec_mom}
		\mu'_i = \mu_i + \mu_j
	\end{equation}
	This moment feels an effective transverse field
	\begin{equation}
		\label{eq:tfim_rule_bonddec_field}
		h'_i = \frac{h_i h_j}{2 J_{ij}}
	\end{equation}
	and has couplings to all of the neighbors of either of the original spins
	\begin{equation}
		\label{eq:tfim_rule_bonddec_coup}
		J'_{ik} = J_{ik} + J_{jk}
	\end{equation}
	In terms of logarithmic variables:
	\begin{eqnarray}
		\beta'_i &=& \beta_i + \beta_j - \zeta_{ij} \\
		\zeta'_{ik} &=& \min\left\{ \zeta_{ik}, \zeta_{jk}\right\}
	\end{eqnarray}
\end{itemize}

In the planar approximation, only the first rule needs to be modified; bond decimation simply collapses two sites joined by an edge and does not generate  non-planar bonds. When a field pins a site with $n$ neighbors, however, all of the $\binom{n}{2}$ possible bonds are in principle generated. In order to maintain local planarity, we therefore add these bonds in from strongest to weakest only when they would not cross a previously added bond. We note that Motrunich \emph{et al} also threw out bonds that were generated by the RG rules by ignoring all bonds weaker than a certain threshold. This does not modify the results of the RG above the scale of the threshold but prevents an explosion in memory requirements.

\subsubsection{Results} % (fold)
\label{ssub:tfim_results}

In the numerical SDRG approach, one creates a bare two dimensional lattice $G$ of some size $L\times L$ with randomly sampled couplings $\beta$ and $\zeta$ from some $P_0(\beta), P_0(\zeta)$ and then runs the decimation procedure directly while monitoring the flow of geometry and couplings. Any infinite randomness fixed points will have much more complicated joint distributions $P_\infty(G,\beta,\zeta)$ governing their couplings and geometry and thus the numerically observed flow will necessarily exhibit a transient behavior as it approaches the scaling regime. Similarly, at the tail of the process as $N$ becomes very small, we expect to see finite size effects modifying the flow. Thus, in order to identify thermodynamic scaling behavior, we would like to see scaling in the coupling distributions $P_\Gamma(\beta)$, $P_\Gamma(\zeta)$ for as many orders of magnitude in $N$ as possible between these two regimes, independently of the initial size $L$. 

\begin{figure}[tbp]
	\centering
		\includegraphics[width=\columnwidth]{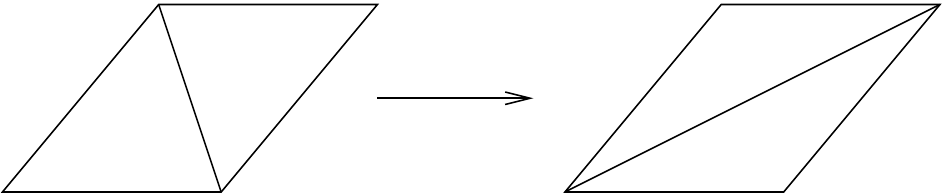}
	\caption{The facet flipping move used to `mangle' initial lattices. By randomly introducing a finite density of these moves, we maintain planarity while broadening the degree distribution toward that obtained at a fixed point.}	
	\label{fig:figs_flip-facet}
\end{figure}

All of our flows begin with periodic triangular lattices of size $L\times L$, which we then `mangle' randomly by applying facet flips to randomly chosen edges as in Figure~\ref{fig:figs_flip-facet}. By applying a finite density of these flips, we broaden the initial degree distribution of $G$ (which for a triangular lattice is simply $P(d) = \delta(d-6)$) and bring it closer to the scaling distribution. Empirically, this suppresses the initial transients observed in the flow. 

\begin{figure}[htbp]
	\centering
	\includegraphics[width=\columnwidth]{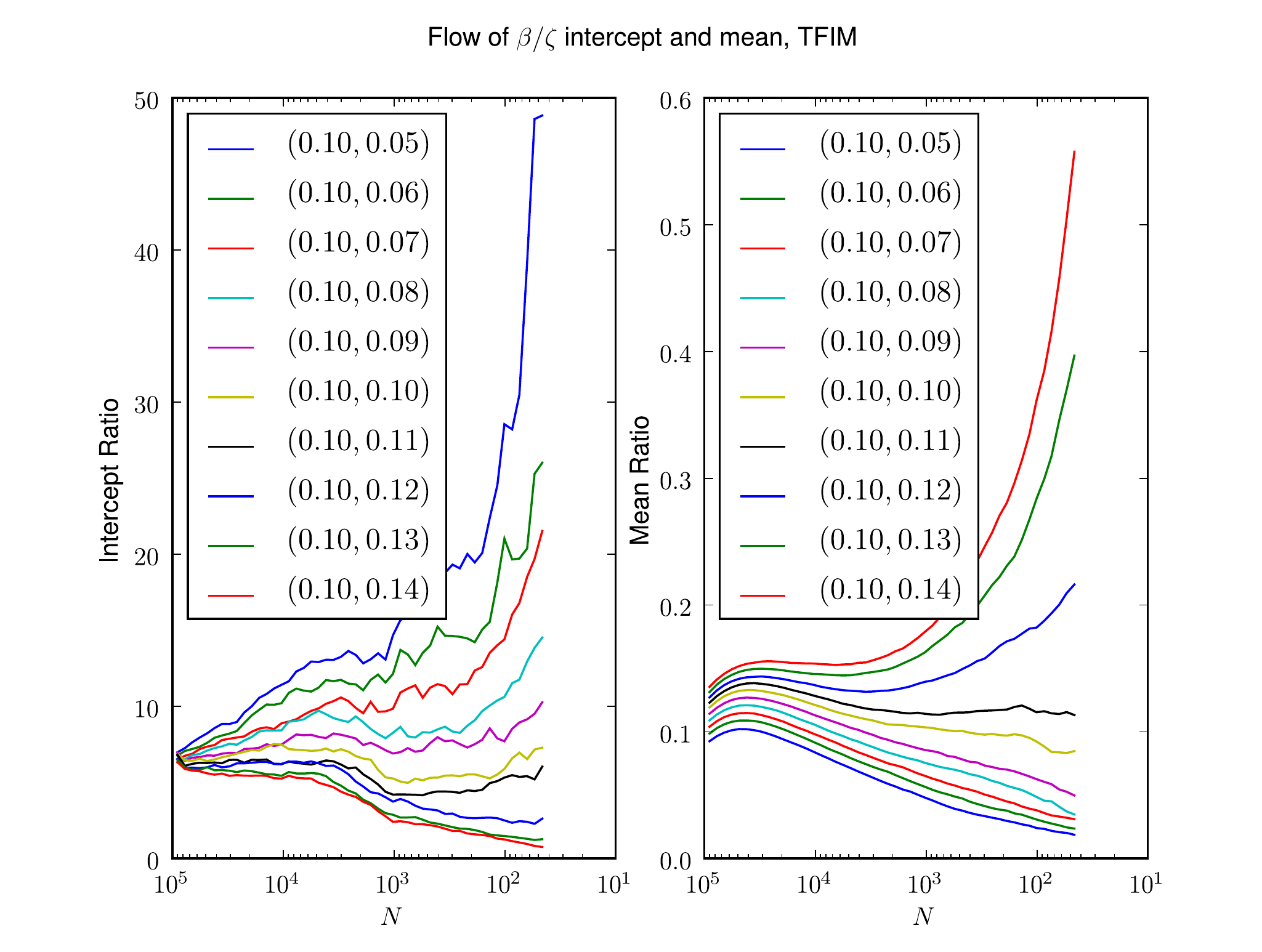}
	\caption{Rough cut through phase diagram near IRCP by varying intercept and slope $(a,m)$ of linear exponential initial conditions for $P(\zeta) = (a + b \zeta)e^{-c \zeta}$. The intercept ratio $R_0 / P_0$ flows to large values in the paramagnet and small values in the ferromagnet and remains finite at the IRCP. The mean ratio $\langle \beta \rangle / \langle \zeta \rangle$ flows the opposite way. We identify our candidate IRCP at $(a,m) \approx (0.10, 0.11)$ by looking for the separatrix in these flow lines.}
	\label{fig:figs-tfim_fig-coarse-a-m-int-meanratio-flow}
\end{figure}

The primary knobs for exploring the phase diagram of the TFIM are the initial distributions $R_0(\beta)$ and $P_0(\zeta)$. As in previous work, we find that the field distribution $R_t(\beta)$ remains approximately exponential throughout all of our flows and thus we always take the initial condition 
\begin{equation}
	R_0(\beta) = e^{-\beta}.
\end{equation}
We also find that the bond distribution $P_t(\zeta)$ tends to develop an upward initial slope as it flows towards the infinite randomness critical point but this distribution gets cut off by a roughly exponential tail by the planar approximation. Thus, we take 
\begin{equation}
	P_0(\zeta) \propto (a + b \zeta) e^{-c \zeta}
\end{equation}
such that we can begin with an initial intercept $a$ and slope $m = b - ac$. We locate the critical point by monitoring the intercept ratio $R(0)/P(0)$ and the mean ratio $\langle \beta \rangle / \langle \zeta \rangle$ and try to find fixed points in their flow as a function of initial conditions. See Figure~\ref{fig:figs-tfim_fig-coarse-a-m-int-meanratio-flow}.

\begin{figure}[htbp]
	\centering
		\includegraphics[width=\columnwidth]{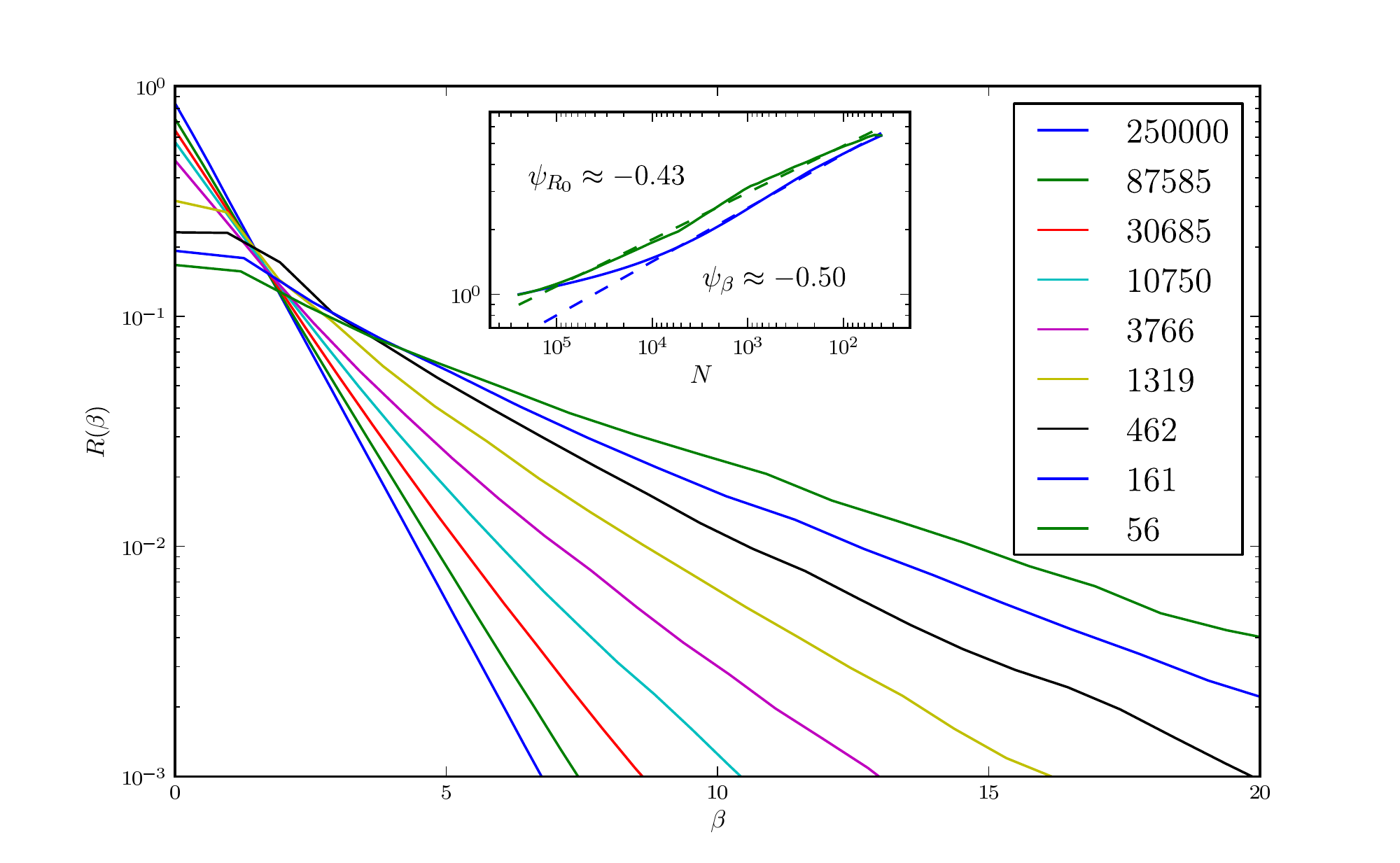}
	\caption{Flow of log field distribution $R(\beta)$ at putative IRCP. Initial lattice $500\times500$, average over 400 runs. (inset) Flow of distribution width as measured by least squares fit to exponential $R(\beta) = R_0 e^{-R_0 \beta}$ and by flow of mean $\langle \beta \rangle$.}
	\label{fig:figs-tfim_fig-ircp-betadist-flow}
\end{figure}

\begin{figure}[htbp]
	\centering
		\includegraphics[width=\columnwidth]{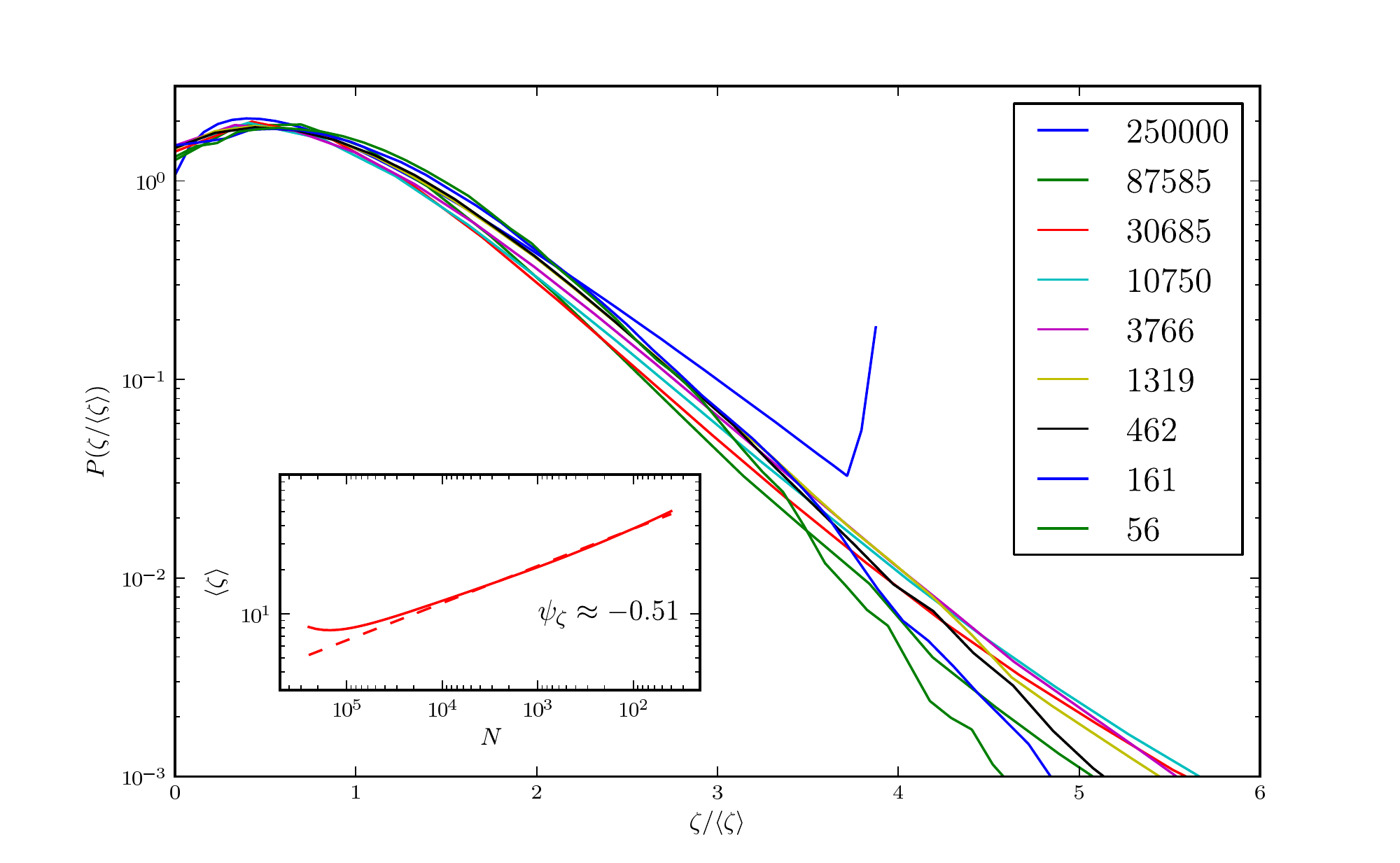}
	\caption{Scaled flow of $P(\zeta / \langle \zeta \rangle)$ at putative IRCP. Same dataset as Fig.~\ref{fig:figs-tfim_fig-ircp-betadist-flow}. (inset) Flow of width as measured by mean $\langle \zeta \rangle$.}
	\label{fig:figs-tfim_fig-ircp-zetadist-flow}
\end{figure}

\begin{figure}[htbp]
	\centering
		\includegraphics[width=\columnwidth]{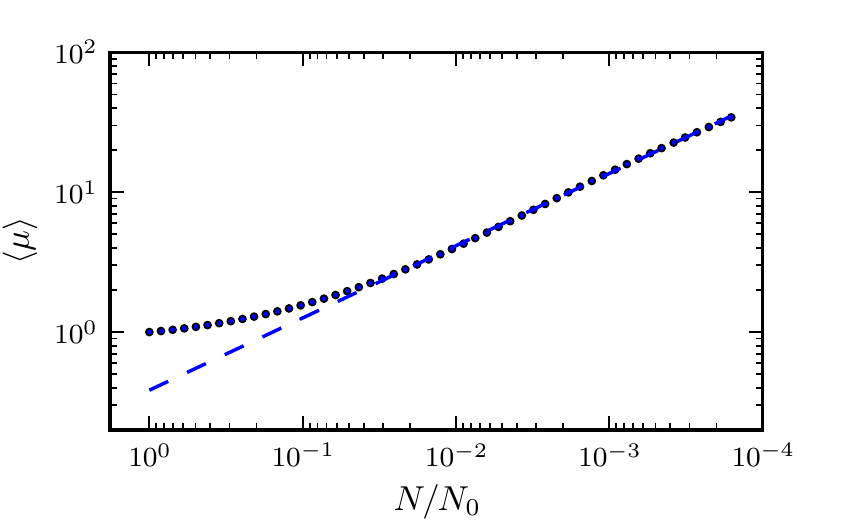}
	\caption{Flow of mean magnetic moment $\langle \mu \rangle$ at putative IRCP. Provides estimate of fractal dimension $d_f \approx 1.03$ from $\langle \mu \rangle \sim N^{-d_f/d}$.}
	\label{fig:figs-tfim_fig-ircp-muflow}
\end{figure}

The primary features of the disordered TFIM in the planar approximation are consistent with those found by Motrunich {\it et al.}:\cite{Motrunich:2000qf}
\begin{enumerate}
	\item The model flows to a random ferromagnet, paramagnet or infinite randomness critical point depending on the relative strength of the initial bond and field distributions. The distributions $R(\beta)$ and $P(\zeta)$ roughly scale at the IRCP. See Figs.~\ref{fig:figs-tfim_fig-ircp-betadist-flow} and \ref{fig:figs-tfim_fig-ircp-zetadist-flow}. 
	
	\item The infinite randomness critical point has a critical exponent $\psi$ defined by $w \sim N^{-\psi / d}$ where $w$ is some measure of the width of the coupling distributions. Asymptotically all reasonable width measures should give the same exponent but they may be estimated numerically in various fashions. Using the same technique as Motrunich \emph{et. al.}, (least squares fitting to an exponential distribution for $P(\beta)$), we find	$\psi \approx 0.43$, which agrees with  Motrunich's result $\psi = 0.42 \pm 0.06$ and also with Monte Carlo and more recent numerical refinements of the SDRG calculation \cite{Pich:1998ba,Kovacs:2010bf}. Alternatively, one may estimate $\psi$ by the flow of $\avg{\zeta}$ (or $\avg{\beta}$), which measures the width of the bond (field) distribution. This provides a higher estimate of $\psi \approx 0.50$ ($0.49$), still consistent with the earlier estimates to within error bars. 
	
	\item The fractal dimension of the spin moments $\mu \sim  N^{-d_f / d}$ may be estimated from the flow of $\avg{\mu}$ with $N$. We find $d_f \approx 1.03$ in agreement with other work. See Fig.~\ref{fig:figs-tfim_fig-ircp-muflow}.
\end{enumerate}

\begin{figure}[htbp]
	\centering
		\includegraphics[width=\columnwidth]{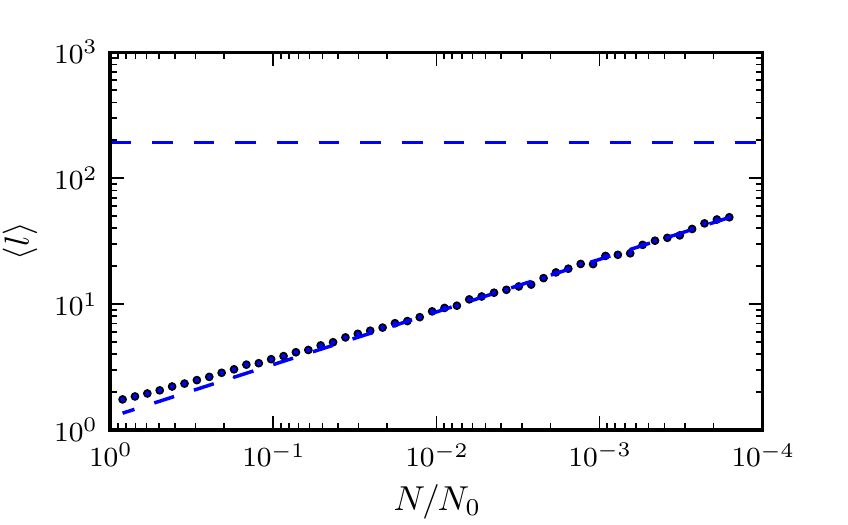}
	\caption{Mean integrated out bond length during IRCP flow. This is a windowed time average of the geometric length of the bonds integrated out in a certain window of the flow. The horizontal line indicates the saturation line expected from a random graph on a torus of size $L=500$; the dashed power law fit gives $\langle l \rangle \sim N^{-0.42}$.}
	\label{fig:figs-tfim_fig-ircp-mblflow}
\end{figure}

The planar SDRG accurately reproduces the phase diagram of the random TFIM as well as the critical exponents of the IRCP as determined by previous numerical studies. This indicates that the truncation of nonplanar bonds is irrelevant to the flow of a system with a strong infinite randomness fixed point.

\subsection{Hardcore Bosons (XX Model)} % (fold)
\label{sub:xx_model}

Disordered quantum XYZ chains have a rich phase diagram studied in considerable detail by Fisher.\cite{Fisher:1994p7999} The existence of infinite randomness fixed points is a hallmark of these one dimensional models. In higher dimensions, Motrunich \emph{et al}\cite{Motrunich:2000qf} did not find a strong numerical flow of the XX model (aka. hard-core bosons) to infinite randomness and did not pursue the model in much detail. 
We confirm this flow toward weaker disorder with a straightforward implementation of the boson SDRG. However, in the planar approximation, we find a spurious but numerically stable marginal fixed point at infinite randomness. This fixed point is characterized by saturation of the lengths of the strong bond and the breakdown of the bare geometry of the system. In this sense, the `fixed point' behavior is not thermodynamic but rather that of a finite size system that has saturated. 

\subsubsection{Model} % (fold)
\label{ssub:xx_model}

The random quantum XX model consists of a lattice of spin 1/2 degrees of freedom $\sigma_i$ governed by the Hamiltonian
\begin{equation}
	\label{eq:xx_ham}
	H = \sum_{\seq{ij}} J_{ij} \left(\sigma_i^x \sigma_j^x + \sigma_i^y \sigma_j^y\right)
\end{equation}
where $J_{ij}$ are random independently distributed couplings. Here we consider only positive couplings $J_{ij}$ and thus the ground state of any single term in the Hamiltonian is a singlet. 

There is only one strong-disorder RG rule for the XX model: 
\begin{itemize}
	\item Bond decimation: the spins $\sigma_i, \sigma_j$ form a singlet and drop out of the effective description of the system. For each pair $k,l$ of their former neighbors, perturbation theory generates an effective interaction by virtual excitations of this singlet:
	\begin{equation}
		\label{eq:xx_rule_bonddec}
		J'_{kl} = J_{kl} + \frac{J_{l}J_{k}}{J_{ij}}
	\end{equation}
	where $J_l$ $(J_k)$ is the largest coupling between $l$ $(k)$ and either of the pair $i,j$. In log variables,
	\begin{align}
	\zeta'_{kl} = \min\left\{ \zeta_{kl}, \zeta_{k} + \zeta_{l} - \zeta_{ij}\right\}.
	\end{align}
\end{itemize}
The flow to an infinite randomness critical point of the TFIM arises when field and bond decimations exactly balance throughout the RG flow -- with only one RG rule, the XX model cannot exhibit such a behavior. Rather, if we begin the scale-free flow with various initial conditions, the only possible behaviors are that the bond distribution gets wider without bound (a standard infinite randomness fixed point), narrower without bound (a finite randomness or pure fixed point) or finds a fixed distribution with width of order the starting width (a marginal infinite randomness fixed point).

% subsubsection model (end)

\subsubsection{Results} % (fold)
\label{ssub:xx_results}

\begin{figure}[htbp]
	\centering
		\includegraphics[width=\columnwidth]{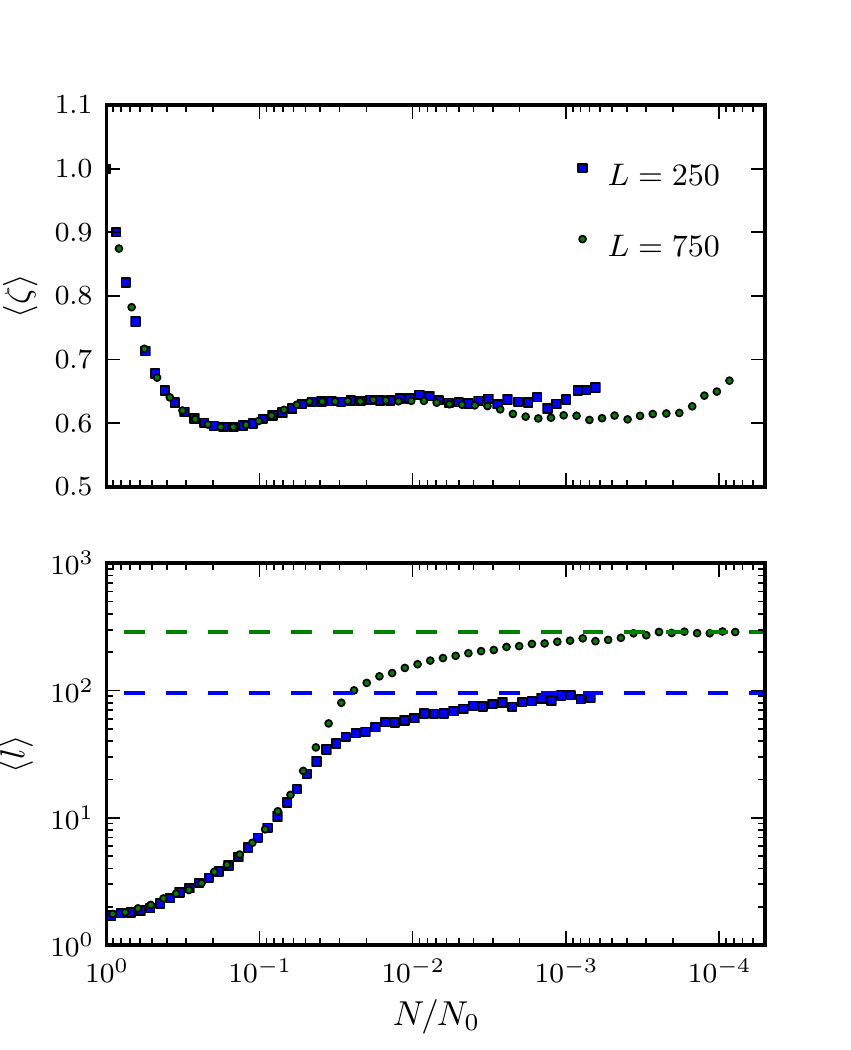}
	\caption{(top) Flow of mean coupling $\zeta = \ln \Omega / J$ and (bottom) mean bond length for hardcore bosons in the planar approximation. Average of 20 planar SDRG runs for each initial lattice size $(250\times250, 750\times750)$, initial conditions as in text. The dashed horizontal lines indicate the expected distance between two randomly chosen points on a torus of size $L$.}
	\label{fig:figs-boson_fig-mbl-twosizes}
\end{figure}

We initialize the planar SDRG with a periodic triangular lattice of size $L\times L$ which we `mangle' randomly (see Sec.~\ref{ssub:tfim_results}) to reduce observed flow transients. The initial couplings $\zeta$ are chosen from an exponential distribution $P_0(\zeta) = e^{-\zeta}$ -- several other initial distributions produced qualitatively similar results. The hardcore boson model flows in three stages (see Fig.~\ref{fig:figs-boson_fig-mbl-twosizes} top): first, the system flows strongly to weaker disorder over about a decade in system size; second, the coupling distribution $P_t(\zeta)$ stabilizes and the system exhibits an apparent marginal infinite randomness plateau; and, third, the flow begins drifting due to finite size effects at the smallest sizes. We have checked that each of these stages of the flow is consistent with the interpretation in terms of initial transient and finite size tail by simply varying the initial system size and comparing the onset and duration of each stage. 

The interpretation of the plateau during the second stage as evidence for a marginal infinite randomness fixed point must be physically spurious, as the system in the absence of the planar truncation exhibits a consistent strong flow to weaker disorder (data not shown).
To understand the plateau better, we consider the geometry of the bare bonds during the flow and find that the mean bond length $\langle l \rangle$ (see Fig.~\ref{fig:figs-boson_fig-mbl-twosizes} bottom) saturates the system size at the end of the first stage of the flow. After this point, the bond length distribution (not shown) is consistent with that of a random graph dropped onto a toroidal geometry. In a physically correct fixed point, these bare lengths should scale with the inverse root of the density of remaining nodes, as they do at the IRCP of the TFIM (Fig.~\ref{fig:figs-tfim_fig-ircp-mblflow}). This `long bond disease' provides a diagnosis of the failure of the RG flow in the Fibonacci case as well.

\subsection{Ising Anyons} % (fold)
\label{sub:maj_model}

Ising anyons are described by an SU(2)  Chern-Simons theory at level $k=2$. They may be equivalently be viewed as neutral Majorana fermion zero modes and their pairwise interactions may be recast as a free fermion hopping problem for real fermions. The undisordered Majorana chain describes a quantum critical point of a spinless superconductor \cite{Kitaev:2000p5909} while the disordered model has the same strong-disorder rules as for hard-core boson hopping and therefore exhibits the same infinite randomness behavior as for the XX chain.\cite{Fisher:1994p7999}  The two-dimensional bipartite case is equivalent to the bipartite imaginary random hopping problem which is known to have a marginally stable infinite randomness phase \cite{Motrunich:2002hz} by various analytic mappings -- this marginal flow is, however, hard to observe in numerical SDRG investigations. We study this model on bipartite and non-bipartite lattices with and without the planar approximation.

\subsubsection{Model} % (fold)
\label{ssub:model_maj}

The Ising anyon model consists of a lattice of pinned anyonic degrees of freedom whose Hilbert space can be built up from the fusion rule
\begin{equation}
	\label{eq:maj_fusion}
	\sigma \otimes \sigma = \mathbf{1}\oplus \psi
\end{equation}
where $\sigma$ represents a Ising anyon and $\mathbf{1}, \psi$ both  represent vacuum states (for our immediate purposes). The Hamiltonian that governs the model is then a sum of pairwise interactions that project a given pair onto one of its two fusion channels. Writing $\Pi^\psi_{ij}$ for the projector onto the $\psi$ channel of fusion of the $i,j$ pair of underlying particles, the Hamiltonian is
\begin{equation}
	\label{eq:maj_ham}
	H = \sum J_{ij} \Pi^\psi_{ij}
\end{equation}
Since the $\mathbf{1}$ and $\psi$ fusion channels are both singlets which act trivially when fused with additional $\sigma$ particles, the sign of the interaction $J_{ij}$ is unimportant to the strong-disorder flow.

There is only one strong-disorder RG rule for the Ising model:
\begin{itemize}
	\item Bond decimation: the anyons $\sigma_i, \sigma_j$ form a singlet and drop out of the effective description of the system. For each neighbor $k$ of $i$ and $l$ of $j$, perturbation theory generates an effective interaction by virtual excitations of this singlet:
	\begin{equation}
		\label{eq:maj_rule_bonddec}
		J'_{kl} = J_{kl} + \frac{J_{l}J_{k}}{J_{ij}}
	\end{equation}
\end{itemize}
The rule is identical to that of the hard-core boson model except that bonds are only generated between next neighbors on opposite (local) sublattices. In particular, if the system begins with a bipartite lattice, the lattice remains bipartite throughout the flow and thus there are two distinct cases for the Ising anyon RG: bipartite and non-bipartite. 
% subsubsection model_maj (end)

\subsubsection{Results} % (fold)
\label{ssub:maj_results}

\begin{figure}[tbp]
	\centering
		\includegraphics{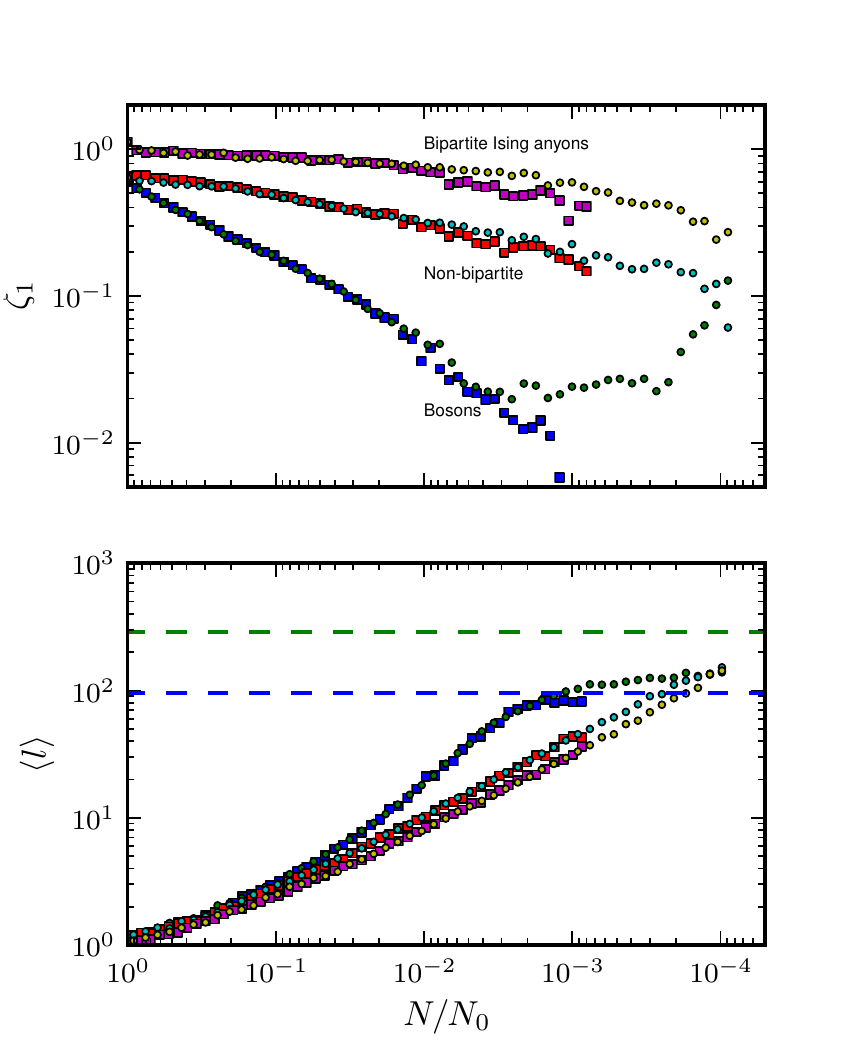}
	\caption{(top) Representative flow of width as measured by median $\zeta_1$ (1 bond/site has strength greater than $\zeta_1$) for bosons, non-bipartite Ising anyons, and bipartite Ising anyons without the planar approximation. Averaged over 50 samples for each size (squares, $L=250$, circles, $L=750$) and model. (bottom) Mean integrated out bond length for same flows, showing saturation in the bosons.}
	\label{fig:figs-npboson_fig-npmix-zeta-bl-twosizes}
\end{figure}

As expected, in the absence of the planar approximation, we find that both bipartite and non-bipartite Ising anyon models flow slowly to weaker disorder, independent of initial conditions. No universal behavior can be extracted, but we indicate typical flows of the width as measured by the median $\zeta_1$ such that 1 bond/site has strength greater than $\zeta_1$ in Fig.~\ref{fig:figs-npboson_fig-npmix-zeta-bl-twosizes}. These flows begin with either triangular (non-bipartite) or square (bipartite) lattices with couplings $\zeta$ drawn from a linear-exponential distribution $P_0(\zeta) = (a + b \zeta)e^{- c \zeta}$ with intercept 1, slope at intercept 2. 
The shallow weakening of the bipartite Ising anyon distribution is consistent with the result that the actual thermodynamic flow is to a marginally stable IRFP,\cite{Motrunich:2002hz} which is too weakly attractive to be found at the sizes we consider. The non-bipartite flow is nearly as shallow, which is suggestive that a similar behavior holds for the case of non-bipartite Ising anyons. 
This is not quite true, however, as evinced by mappings of the non-bipartite pinned Ising anyon problem into the disordered free fermion problem in symmetry class D, where a disordered gapless phase exists,\cite{Laumann:2011vn} but which does not exhibit infinite randomness scaling.

Within the planar approximation, both bipartite and non-bipartite Ising anyon
models `fall apart' at a finite time (decimation fraction) in the RG flow. This
arises due to the extreme truncation imposed by the opposite sublattice rule and
the planar approximation -- a finite local sequence of decimations can leave
behind disconnected sites. Thus, at a finite decimation fraction (of roughly
$1.7\%$, measured for initial sizes $L=300, 500$, averaged over 40 runs each),
such motifs reduce the system to a collection of disconnected sites and the RG
stops prematurely.

\subsection{Fibonacci Anyons} % (fold)
\label{sub:fibonacci_anyons}

\begin{figure*}[t]
	\centering
		\includegraphics{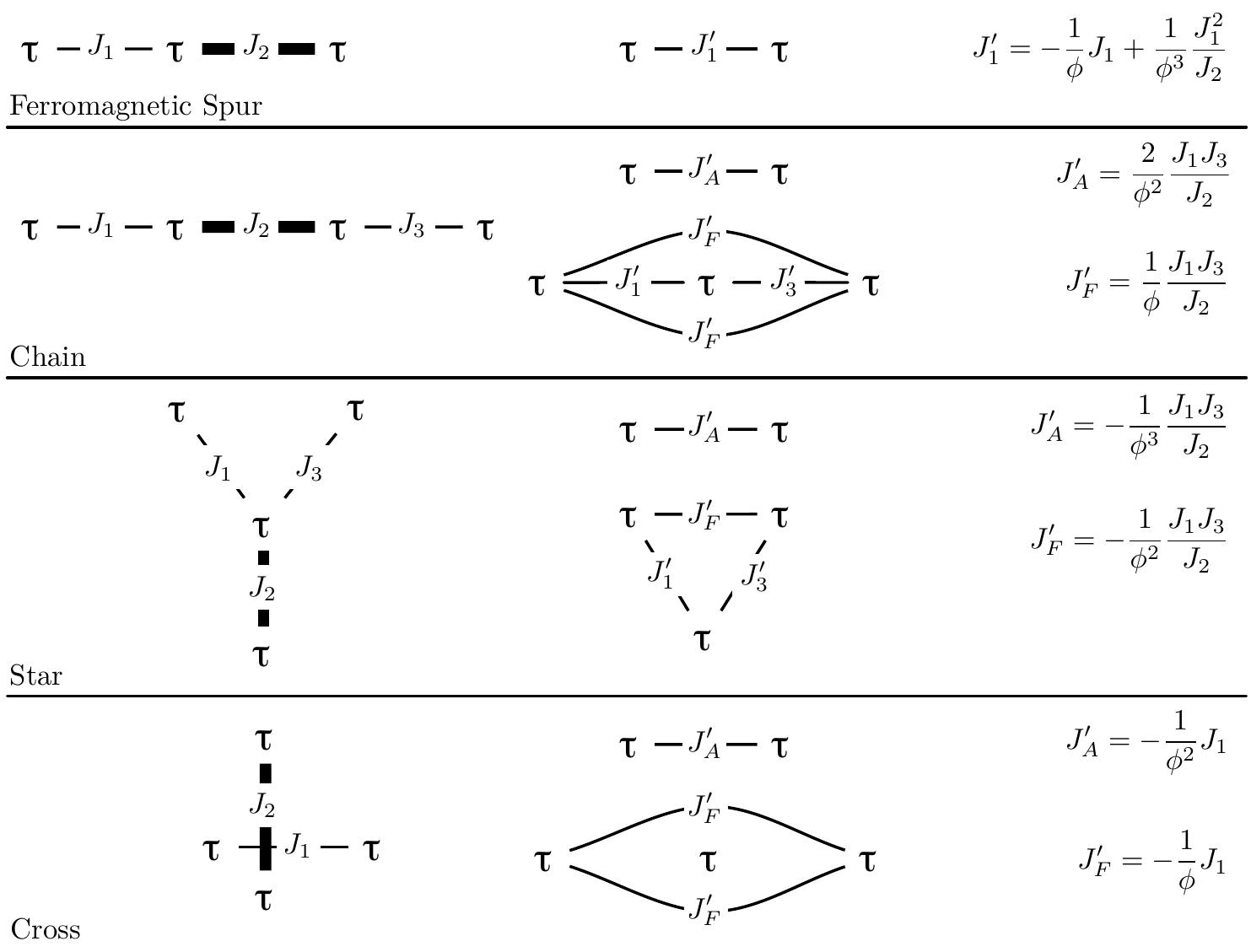}
	\caption{Strong-disorder rules for Fibonacci anyon models. In each configuration on the left, the thick bond $J_2$ indicates the strong bond to be integrated out. For the chain, star and cross configurations, a strong (anti-)ferromagnetic bond leads to the configuration in the lower (upper) row with renormalized coupling $J'_F$ ($J'_A$). Positive couplings are antiferromagnetic (favor the vacuum channel). In the infinite disorder limit, the factor of the golden mean $\phi = \frac{1+\sqrt{5}}{2}$ may be dropped, but the signs remain relevant.}
	\label{fig:figs-tau_anyon-rules}
\end{figure*}

The Fibonacci anyons are described by an SU(2) Chern-Simons theory at level $k=3$. They have attracted some interest in the context of topological quantum computing proposals \cite{Nayak:2008dp} since they are the simplest non-Abelian anyons whose braiding rules are universal for quantum computation.\cite{Freedman:2002} They are also candidate quasiparticles for the $\nu=12/5$ fractional quantum Hall phase.\cite{Rezayi:2009kv,Read:1999wx} The Fibonacci chain, \emph{aka.} the golden chain, was the first anyon lattice model to be studied in detail,\cite{PhysRevLett.98.160409,PhysRevB.78.224204} both in its pure and disordered form, and exhibits an infinite randomness phase. In two dimensions, this is the simplest model which suffers from the second difficulty described in Sec.~\ref{sec:strong_disorder_rg}: fusion of a pair of anyons may energetically influence the fusion of a disjoint pair of anyons elsewhere in the system (see the `cross' configuration in the rules, Fig.~\ref{fig:figs-tau_anyon-rules}). Thus, this is the  model which exhibits the full topological non-locality that motivated the use of the \emph{planar} approximation in our treatment.

\subsubsection{Model} % (fold)
\label{ssub:tau_model}

The Fibonacci anyon model consists of a lattice of pinned anyonic degrees of freedom whose Hilbert space can be built up from the fusion rule
\begin{equation}
	\label{eq:fib_fusion}
	\tau \times \tau = \mathbf{1} + \tau
\end{equation}
where $\tau$ represents a Fibonacci anyon and 1 represents the singlet or vacuum state. The Hamiltonian that governs the model is then a sum of pairwise interactions that project a given pair onto one of its two fusion channels. Writing $\Pi^\tau_{ij}$ for the projector onto the $\tau$ channel of fusion of the $i,j$ pair of underlying particles, the Hamiltonian is
\begin{equation}
	\label{eq:fib_ham}
	H = \sum J_{ij} \Pi^\tau_{ij}
\end{equation}
Thus a positive coupling $J_{ij}$ corresponds to an ``antiferromagnetic'' term whose ground state is the ``singlet'' $\mathbf{1}$ fusion channel. Restricted to a planar lattice of interactions, this is a sufficient description of the Hamiltonian. In general, however, the pair-wise interaction of two distant anyons should also be labeled by a path connecting the two anyons which indicates on which side of the other anyons in the system the interaction is mediated. 

The appropriate strong-disorder RG rules can be derived from an application of perturbation theory to various clusters of $\tau$ particles. Unlike in the TFIM and XX models, the sign of the interaction is critically important to the physical behavior: a strongly coupled pair of $\tau$s can either fuse to  a single new $\tau$ or to the vacuum $\mathbf{1}$ state. In both cases new neighbor interactions are generated but with signs that depend on the topology of the local interaction network. Moreover, crossed bonds that involve otherwise non-interacting $\tau$'s interact under renormalization due to the non-locality of the Hilbert space; in particular, decimation of a crossed bond flips the sign of the remaining interaction and/or generates multiple path dependent interactions. The rules are summarized in Fig.~\ref{fig:figs-tau_anyon-rules}.

\subsubsection{Results} % (fold)
\label{ssub:tau_results}

As in the other models, we initialize the Fibonacci anyon planar SDRG with a periodic triangular lattices of size $L\times L$, which we `mangle' with a finite density of facet flips (see Sec.~\ref{ssub:tfim_results}). We choose independent random couplings $\zeta$ for each bond from an exponential or linear-exponential distribution $P_0(\zeta) \propto (a+b\zeta)e^{-c \zeta}$. We tune these initial conditions in order to minimize the transient in the flow behavior and see as many decades as possible of scaling behavior.

\begin{figure}[htbp]
	% from results/tau
	\centering
	\includegraphics[width=\columnwidth]{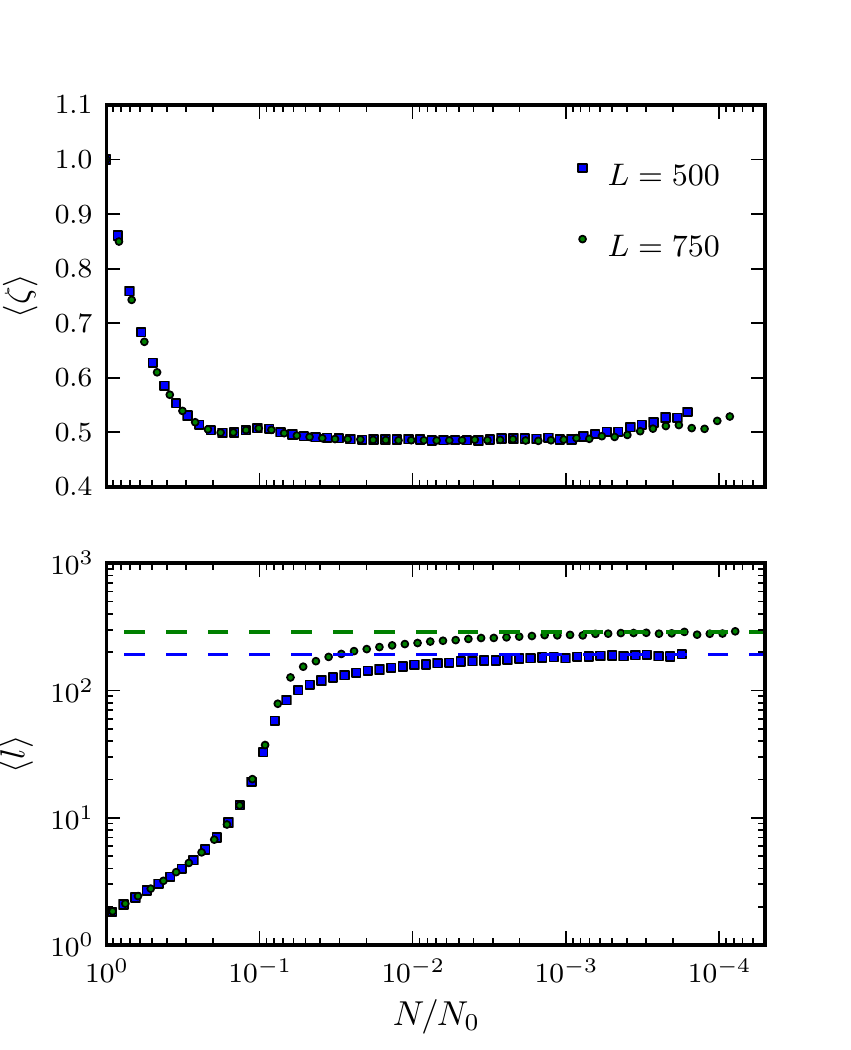}
	\caption{(top) The flow of the coupling width $\langle \zeta \rangle$ downward to a plateau is falsely suggestive of a marginal infinite randomness fixed point. (bottom) The saturation of the mean decimated bond length $\langle l \rangle$ reflects the pathology of this fixed point, which has lost all sense of the two-dimensionality of the initial system. Average of 50 planar SDRG runs for each initial lattice size $(500\times500, 750\times750)$, initial conditions as in text. The dashed horizontal lines indicate the expected distance between two randomly chosen points on a torus of size $L$.}
	\label{fig:figs_tauresults-mbl-flow-two-sizes-rel}
\end{figure}

Under the planar SDRG, the Fibonacci anyons exhibit a spurious flow to a stable fixed point in which the coupling distribution is essentially fixed (full distribution not shown). See Figure~\ref{fig:figs_tauresults-mbl-flow-two-sizes-rel}. The plateau in the coupling width after an initial transient is a robust feature of the planar flow, independent of the initial size and therefore suggestive of thermodynamic behavior -- just as the upturn at small sizes shifts with the initial size, which agrees with the interpretation of this upturn as a finite size effect. Unfortunately, the underlying geometry of this fixed point reflects the `long bond disease' also exhibited by the boson flow and inconsistent with a physical fixed point (see Sec.~\ref{ssub:xx_results}). As can be seen in Fig.~\ref{fig:figs_tauresults-mbl-flow-two-sizes-rel}, the mean bond length $\langle l \rangle$ of the strongest bond in the system saturates to the bare system size at essentially the same point in the flow that the coupling width plateaus. 

\section{Implementation} % (fold)
\label{sec:implementation}

The primary obstacle to implementing a \emph{planar} strong-disorder renormalization group is the maintenance of the planarity condition on the underlying interaction graph. The usual representations of the interaction graph by either adjacency matrix or edge lists does not contain information about the embedding of the graph into a two dimensional manifold, which quickly becomes impossible to discern. Similarly, using a geometric embedding into a planar or toroidal surface is problematic because the bare geometry of the interaction graph may become very complicated during the renormalization flow. Rather, in order to maintain the local planarity of the interaction graph, we view it as the edge mesh on a topologically oriented surface represented using a halfedge data structure, familiar in the computational geometry literature. In this structure, interactions between vertices are represented by pairs of halfedges which define the orientation of the adjacent facets on the surface (see Fig. 1 of Ref.~\onlinecite{Bronnimann:2001ys}). As the interactions and sites are decimated, we update this data structure in a topology preserving manner and thus guarantee that we always maintain local planarity correctly.

We have implemented our own halfedge data structure library in scientific Python \cite{Jones:2001zr} and run all planar renormalization groups using this code. The models studied without the planar approximation were also implemented in Python using the significantly simpler representation provided by adjacency lists.

% section implementation (end)

\section{Conclusions} % (fold)
\label{sec:conclusion}

In summary, we have introduced the \emph{planar} approximation into the strong-disorder RG treatment of two dimensional pinned anyons in order to handle the crossing interactions generated by the Fibonacci anyon RG rules. By comparing the SDRG flow with and without the planar approximation on a variety of models, we find that the approximation accurately reproduces physics near systems flowing strongly to infinite randomness, such as at the random TFIM critical point, but breaks down in one of several characteristic ways in systems flowing back to weaker disorder. The similar pathologies of the hard-core bosons and the Fibonacci anyons strongly suggest that the Fibonacci model has no infinite randomness fixed points.

If a method can be developed to treat the topologically non-local interactions of the Fibonacci model without a planar truncation, it is conceivable that it would reverse our conjectured flow to weaker disorder. 
We believe this is unlikely and that the similarity to that of the 2-D hardcore bosons will continue to hold: the Fibonacci anyons do not exhibit any infinite randomness physics. 
However, this does not rule out the possibility of other, nontrivial disorder-induced gapless phases in the Fibonacci phase diagram, as have been recently established in the form of a thermal metal phase for disordered pinned Ising anyons.\cite{Laumann:2011vn} 

% section conclusion (end)

\section{Acknowledgments} % (fold)
\label{sec:acknowledgments}

C.R.L. acknowledges support from a Lawrence Gollub fellowship and the NSF through a grant for the Institute for Theoretical Atomic and Molecular Physics (ITAMP) at Harvard University. D.A.H. was supported, in part, by NSF DMR-0819860. A.W.W.L. was supported, in part, by NSF DMR-0706140. G.R. was supported, in part, by the Packard Foundation and the IQIM, and an NSF PFC with
support of the Moore Foundation.

% section acknowledgments (end)

\bibliography{disorder}

\end{document}